# Soliciting Stakeholders' Fairness Notions in Child Maltreatment Predictive Systems

Hao-Fei Cheng
University of Minnesota
cheng635@umn.edu

Logan Stapleton
University of Minnesota
stapl158@umn.edu

Ruiqi Wang
Carnegie Mellon University
ruiqiw1@andrew.cmu.edu

Paige Bullock
Kenyon College
bullock1@kenyon.edu

Alexandra Chouldechova
Carnegie Mellon University
achould@cmu.edu

Zhiwei Steven Wu
Carnegie Mellon University
zstevenwu@cmu.edu

Haiyi Zhu
Carnegie Mellon University
haiyiz@cs.cmu.edu

## ABSTRACT
Recent work in fair machine learning has proposed dozens of technical definitions of algorithmic fairness and methods for enforcing these definitions. However, we still lack an understanding of how to develop machine learning systems with fairness criteria that reflect relevant stakeholders' nuanced viewpoints in real-world contexts. To address this gap, we propose a framework for eliciting stakeholders' subjective fairness notions. Combining a user interface that allows stakeholders to examine the data and the algorithm's predictions with an interview protocol to probe stakeholders' thoughts while they are interacting with the interface, we can identify stakeholders' fairness beliefs and principles. We conduct a user study to evaluate our framework in the setting of a child maltreatment predictive system. Our evaluations show that the framework allows stakeholders to comprehensively convey their fairness viewpoints. We also discuss how our results can inform the design of predictive systems.

## 1 INTRODUCTION
Machine learning (ML) algorithms are increasingly being used to support human decision-making in high-stakes contexts such as online information curation, resume screening, mortgage lending, police surveillance, public resource allocation, and pretrial detention. However, concerns have been raised that algorithmic systems might inherit human biases from historical data, and thereby perpetuate discrimination against already vulnerable subgroups. These concerns have given rise to a rapidly growing research area of fair machine learning. Recent work in this area has produced dozens of quantitative notions of algorithmic fairness [2, 17, 24, 30, 57, 71], and provided methods for enforcing these notions [1, 2, 24, 41, 42, 80].

Existing research on fair machine learning has primarily focused on fairness at the level of pre-defined groups. This *group fairness* approach first fixes a small collection of groups defined by protected attributes (e.g., race or gender) and then asks for approximate equality of some statistic of the predictor, such as positive classification rate or false positive rate,[1] across these groups (see, e.g., [1, 30, 46]). While notions of group fairness are easy to operationalize, they are aggregate in nature and make no promises of fairness to finer subgroups or individuals [24, 31, 42]. In contrast, the *individual fairness* approach aims to address this limitation by asking for explicit fairness criteria at an individual level. For example, Dwork et al. [24] propose an individual fairness notion that requires that similar people are treated similarly. Their formulation of fairness crucially relies on a task-specific metric that captures whether two individuals are similar for the purpose of the task at hand. Due to the challenges of specifying such a metric in any given real-world decision-making problem, it remains difficult to operationalize individual fairness in practice.

Irrespective of the approach one takes to quantify fairness, it is important to engage relevant stakeholders in the design of real-world decision-making systems. As Shah [66] has argued, achieving legitimacy or "social license" from the broader community is critical to the ability of even the best-conceived technologies to have a positive social impact. Similarly, [47, 77] recommend that *stakeholders* affected by the decisions should be centered in these processes.

One example of a "fair" technology that failed to be adopted due to a lack of stakeholder support is a school start time scheduling tool proposed in Boston intended to decrease bussing costs while improving racial equity and better accommodating differences in circadian rhythms across students of different ages. The system's design failed to account for the excess burden that the proposed times would place on families with multiple children who attend different schools, particularly for lower-income parents who tend to have inflexible work schedules [76]. This is not an isolated example. In a recent study, Veale et al. [70] interviewed 27 public sector ML practitioners across 5 OECD countries and noted the common disconnects between current fair ML approaches and the organizational and institutional realities, constraints, and needs in which algorithms are applied.

---

[1] *False positives* occur when a subject has true negative label, but a classifier erroneously classifies the subject positively. For example, if a child is truly at low risk of maltreatment, but a classify predicts that they are a high risk, this is a false positive. *False negatives* occur when subjects with true positive labels are negatively classified.

Thus, involving affected stakeholders in the algorithm design process—particularly, in the process of defining fairness—is of utmost importance. To this end, we propose a novel framework for eliciting stakeholders' opinions around algorithmic fairness. The framework combines two components: an **interactive interface** that allows stakeholders to examine the data and audit an algorithm's predictions, and an **interview protocol** that is designed to probe stakeholders' thoughts and beliefs on fairness and biases of the algorithm while they are interacting with the interface.

We evaluated our framework in the high-stakes context of developing machine learning-based risk assessment tools to assist child abuse hotline call workers in their screening decisions. Our work is motivated by the Allegheny Family Screening Tool (AFST), which has been used in Allegheny County, PA since the summer of 2016 [69]. We conducted in-depth interviews with 12 participants from two groups of stakeholders (parents and social workers) to understand their fairness viewpoints. The interviews allow us identify fairness approaches that align with stakeholders' beliefs, and allow stakeholders to provide rich reasoning to explain their viewpoints. For child maltreatment risk assessment, the stakeholders we interviewed slightly preferred equalized odds (i.e., equalizing accuracy at identifying low- and high-risk cases across the sensitive attributes), compared to unawareness (i.e., not considering the sensitive attributes at all) and statistical parity (i.e., equalizing high-risk predictions across sensitive attributes). When asked to make individual fairness comparisons, there was little agreement between these stakeholders in most scenarios.

We propose a novel method for engaging human stakeholders in the algorithm design process - in the process of defining fairness. Our work also contributes empirical understanding of stakeholders' fairness opinions in the high-stakes context of developing machine learning-based risk assessment tools.

## 2 RELATED WORK
### 2.1 Fairness in Machine Learning

There has been significant development in research on machine learning fairness and accountability in recent years [2, 17, 30, 57, 60, 71]. Prior literature on ML fairness can generally be classified in two categories: group fairness and individual fairness. The more commonly studied notion, group fairness, requires parity of some statistical measure across a fixed number of protected groups. In this paper, we ask study participants about three of the most popular notions of group fairness: *fairness through unawareness* (henceforth *unawareness*), which is the notion that in order to be fair, an algorithm should explicitly not consider a protected attribute (e.g. race or gender) when making its decisions [59]; *statistical (or demographic) parity*, which entails that a fair algorithm have parity of positive classification rates across a fixed number of protected groups; and *equalized odds* [30], which entails that a fair algorithm have equal accuracies —true positive and false positive rates— across a fixed number of protected groups. While all three notions offer some theoretical fairness guarantees, they also have different shortcomings. First, unawareness has long been critiqued: [59] argue that even when protected attributes are not considered as predictive features, there may be "background knowledge" —other data which serves as proxy or strong predictor for the removed attributes— which recreates the effect of including the removed attributes, e.g. someone's zipcode may be a strong predictor of their race. In general, policy and decision making which actively disregards sensitive attributes, e.g. color-blind policies to mitigate racial discrimination, have been critiqued, as well [4]. Second, [30] critique statistical parity on the grounds that 1) it is not fair insofar as it "permits that we accept the qualified applicants in one demographic, but random individuals in another, so long as the percentages of acceptance match"; and 2) we might incorrectly classify a number of samples in order to maintain equal rates of positive classification. Third, equalized odds can be impossible to achieve simultaneously with other common fairness notions, like statistical parity or calibration [16, 46]. In general, group fairness metrics provide no meaningful guarantees of fairness to individuals or more refined sub-groups [24, 31, 42].

On the other hand, notions of individual fairness explicitly constrain algorithmic decisions at an individual level [24, 37]. For example, the individual fairness notion in [24] requires "treating similar individuals similarly;" the meritocratic fairness notion in [37] requires that the algorithm should "prioritize more deserving individuals." However, these approaches require strong assumptions, such as a consistent measure for similarity or merits across individuals, which usually do not hold in real-world contexts. Furthermore, [77] suggests centering research not on proposing new technical definitions, but rather on proposing new procedures for involving stakeholders to determine which notion of fairness is best. To this end, recent work [8, 33, 38] provides theoretical models of human auditors or arbiters who can provide fairness feedback to assist an algorithm to provably enforce individual fairness without an explicit similarity measure. Beyond individual fairness, there has also been work that involves human efforts in developing algorithms [5, 29, 51, 81], making final decisions after algorithm recommendations [48], and making decisions about fairness trade-offs [79] (as satisfying the criteria for all fairness definitions is mathematically impossible [26, 45]). However, some critique previous elicitation methods for not capturing the *reasons* behind responses [34, 77]. Additionally, it remains a major challenge to devise mechanisms to involve stakeholders in algorithmic development and auditing that do not require unrealistic levels of technical knowledge among participants.

### 2.2 HCI Research on Algorithmic Fairness

With the increased attention on algorithmic fairness, researchers and practitioners designed novel visualization techniques to help people examine the machine learning algorithms and identify biases. For examples, the What-if Tool and AI Fairness 360 are two open-source tools that allow users to visually examine the behavior of their machine learning models and identify potential biases [9, 75]. Other similar visual analytics system are also developed that allow users to audit the group and subgroup fairness of machine learning models [13], or to help data scientists and practitioners make fair decisions [3].

More recently, HCI researchers have begun to investigate human perspectives on algorithmic fairness. Several recent studies have investigated public [28, 64, 65, 72, 74] and practitioner [32, 56, 70] perspectives on the use of algorithmic systems for public-sector decisions. This body of work suggests that fairness principles need to be context-specific, and the algorithmic systems should embody the fairness notions derived from the community of stakeholders [10, 11, 23, 49, 50, 63]. There has been encouraging work towards this direction. For example, researchers have conducted workshops and interviews to understand what people think fairness means in the context of resource allocation [52] or targeted online ads [78]. Researchers have also conducted surveys to gauge well non-technical subjects understand existing fairness metrics [63], how explaining these fairness differently affects subjects' beliefs about fairness [10], and what features should or should not be used by a fair learning algorithm [28, 65]. While these studies provide us with a better understanding of general public and user perceptions of justice and fairness, it is often difficult to translate these qualitative understandings into the system criteria and directly inform the algorithm developments.

### 2.3 Hybrid Approach

More recently, interdisciplinary research teams have begun investigating how hybrid approaches combining the tools of both HCI and machine learning can be effectively applied in developing fair and accountable algorithmic systems. Such work includes the WeBuildAI framework [53] and related approaches [40, 61] for incorporating stakeholder preferences into allocation decisions. A recent workshop on Participatory Approaches to Machine Learning held at ICML 2020 featured recent and ongoing work in this emerging space. Featured research included studies of recommender systems [19], approaches to patient triage during the COVID-19 pandemic [36], and critical assessments of the role of participatory methods in algorithm design [62].

## 3 FRAMEWORK FOR ELICITING FAIRNESS FROM STAKEHOLDERS

### 3.1 Research Question

In the paper, we want to answer the following research question: **How can we effectively elicit fairness notions from a community of stakeholders who are not technical experts?**

To answer this question, we propose an elicitation framework that consists of two components: (1) an ***interactive interface*** that allows stakeholders to express their subjective fairness notions, and (2) an associated ***interview protocol*** that further probes stakeholders' reasoning behind their elicited notions (see Figure 1).

### 3.2 Interface Design

*3.2.1 Design goals.* The goal of the interface is to enable stakeholders to express their perspectives by reasoning about the algorithm's impact at different levels, ranging from individual decisions to the effects on demographic groups.
**Goal 1: Elicitation at the "macro" level.** Corresponding to the "group fairness" notion, the interface should enable users to examine the data and algorithm performance in the groups defined by the users (not limited to groups defined by common protected attributes such as gender and race). The interface should present the various statistical metrics for each subgroup and visualize them for stakeholders to investigate. The stakeholders can then express whether each statistical group fairness measure is aligned with their perspectives.
**Goal 2: Elicitation at the "micro" level.** Corresponding to the "individual fairness" notion, the interface should enable users to inspect the data and algorithm recommendations at a case-by-case level. Combining the approaches from prior work [8, 38, 53], the interface elicits individual fairness feedback by asking stakeholders to make two types of pairwise comparisons: (1) whether the pair of individuals should be treated similarly or not, and (2) whether one individual should be prioritized over the other one or not.
**Goal 3: Elicitation at the "meso" level.** The goal is to enable stakeholders to compare any single selected case with all other cases in the dataset. Different stakeholders may have different criteria for evaluating the similarity and priority across the cases. Thus the interface should allow users to specify their own metrics when exploring the data.

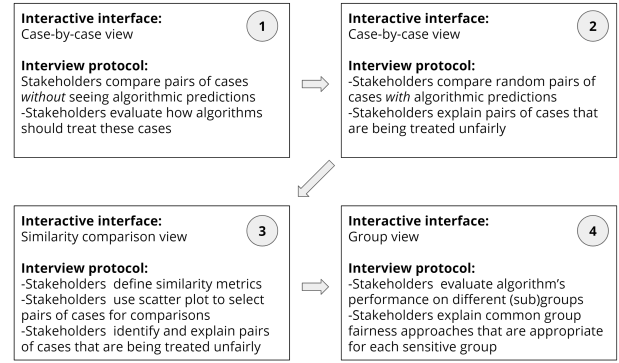

Figure 1: Fairness elicitation framework using our interactive interface and interview protocol. Arrows indicate the progression of our interviews,[2] starting from the case-by-case view (step 1) and proceeding to the group view (step 4).

*3.2.2 Interactive interface.* Our interactive interface prototype[3] consists of three primary views: (i) a group view corresponding to Goal 1 (Figure 2a), (ii) a case-by-case view corresponding to Goal 2 (Figure 2b), and (iii) a similarity comparison view corresponding to Goal 3 (Figure 2c).
**Group view:** This view aims to give users a holistic view of the algorithm's performance by showing how it varies across groups according to different metrics. Users have the option to select from a list of common classification performance metrics. The drop-down menus allow the user to select attributes with which to separate the data into subgroups. The interface displays a bar chart depicting the algorithm's performance across the specified subgroups. A textual description is also provided below the graph to provide an alternate description of the algorithm's performance. The visualization corresponds to group fairness notions and the interactive interface

---

[2]In our study, we changed the order of the interview slightly by asking participants about their viewpoints on common group fairness approaches in step one due to time consideration.

[3]A demo of our interface can be assessed here (note that the data shown in the demo are synthetic): https://z.umn.edu/fairnessElicitationInterface

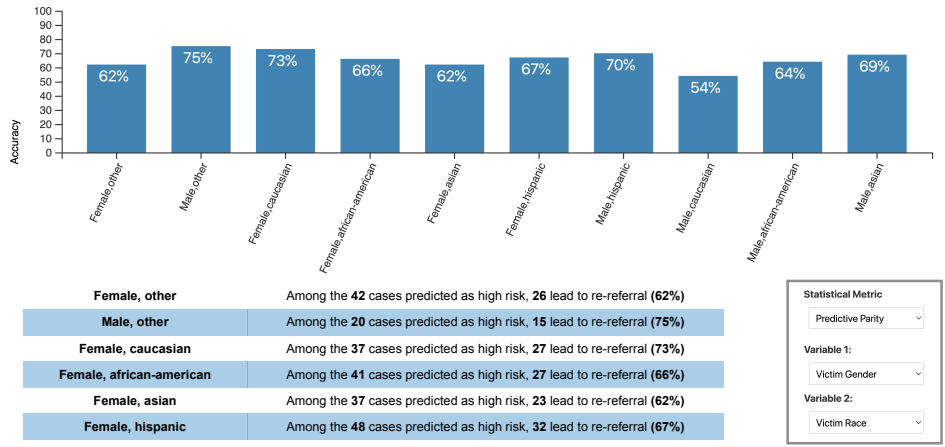

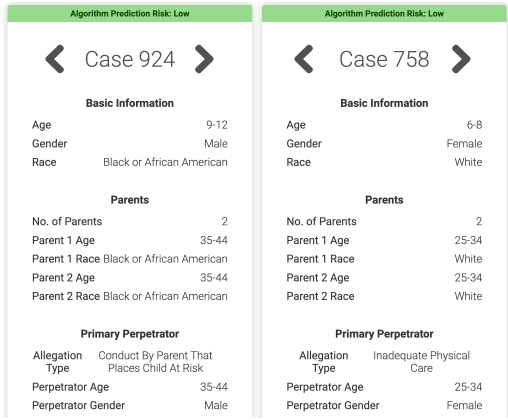 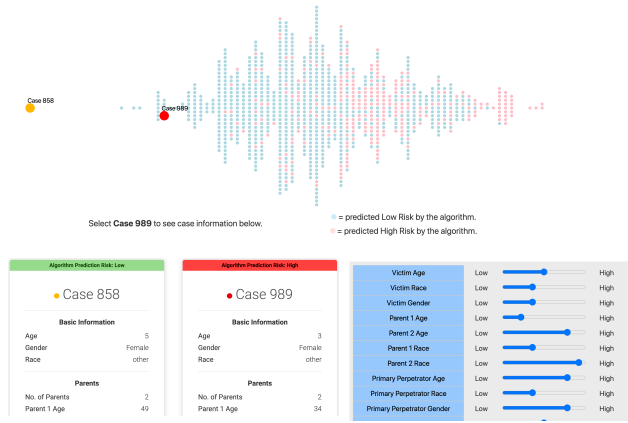

**Figure 2: Our fairness elicitation interface contains three different views, which allows stakeholders examine the algorithm at different levels. In this example, the interface is presenting synthetic data from a child maltreatment prediction system (see Section 4.3).**

allows users to explore any group or performance metrics they are interested in.

**Case-by-case view:** This view allows users to deliberate the algorithm at a granular level of individual predictions. Each case of algorithm prediction is presented as a card; the interface shows two cases at a time for pairwise comparison. On each card, the algorithmic prediction is shown on top, followed by features the algorithm used to make the prediction. Hovering over each feature will show users the detailed description of that feature, and the possible values the feature can take. Users can browse through the cases back and forth. The tool will randomly select a new case from the dataset, and replace the currently displayed case. Users can explore new cases by changing the case on either the left or right. The interface lets stakeholders inspect the profiles and detailed features of any two individuals treated by the algorithm, allowing them to determine if the pair should be treated equally. This aligns with the definition of individual fairness [24] and works that operationalize it (e.g. [8, 39]).

**Similarity comparison view:** This view shows a one-dimensional scatter plot that compares a selected reference case with all other cases in the dataset. This allows users to explore the dataset at a macro view and narrow down to individual cases for inspection. This scatter plot displays all the cases in the dataset, with each case represented by a dot on the plot, color-coded according to the algorithm prediction. The reference case is positioned at the far left of the plot, with other cases ordered by similarity to the reference case along the x-axis. A weighted Euclidean distance metric is used to calculate the similarity of the cases.[4] The y-axis shows the distribution of the cases at that similarity level. A control panel allows users to change the weight associated with each feature. Users can customize the weights to re-rank the cases in an order that aligns with their viewpoints. Users can select a case from the plot to compare with the reference case, or set a new case as the reference case. The similarity comparison view allows stakeholders to compare a reference case with all the other cases. The dots at the same position on the X-axis have the same similarity score (same distance from the reference case), which allows users to quickly see the distribution of the similarity scores across a large number of cases. This allows them to quickly narrow down to individual cases from the whole dataset for comparison, such as looking at cases where the cases with high level of feature similarity and evaluate if they should be receive the same decision.

---

[4]The weighted Euclidean distance between cases $p$ and $q$ is calculated by: $\sqrt{\sum_{i=1}^{n} w_i (q_i - p_i)^2}$, where $w_i$ denotes the user-assigned weight for feature $i$.

### 3.3 Interview Protocol

To complement this interface, we develop interview protocols to probe stakeholders' fairness viewpoints and principles. Our protocols are based on the think-aloud approach, which is one of the most valuable usability engineering methods in HCI [58]. We ask stakeholders to use the interface we described above "while continuously thinking out loud—that is, verbalizing their thoughts as they move through the user interface"[58]. Think aloud serves as "a window on the soul," letting us discover what participants really think about the fairness and bias of the algorithm [58].

First, we ask stakeholders to compare pairs of cases in the data *without showing* the algorithmic predictions, in the case-by-case view (Figure 2b). We ask stakeholders if both cases should be treated equally (i.e. receive the same prediction by the algorithm), and if not, what alternative outcomes should the two cases receive to align with the stakeholders' fairness principles. In this stage, we only show users the features for the cases, as we aim to collect stakeholders' fairness notions regardless of the predictions of those outcomes, and the factors they would consider when evaluating the cases in the context. Participants start by comparing pairs of cases which differ by only one factor, then move onto pairs which differ by two or more factors. At this stage, we also ask our participants whether three common group fairness approaches (unawareness, statistical parity, and equalized odds) are appropriate with respect to sensitive attributes (e.g. for child maltreatment, these are victim age, victim gender, family race, use of public assistance service and perpetrator gender). For a given sensitive attribute, we elicit opinions on whether the following approaches should be met for the algorithm to be fair:

(1) the sensitive attribute should not be a predictive factor (*unawareness*);[5]
(2) the rates of positive classification should be equal across a sensitive attribute (*statistical parity*); or
(3) the false positive and false negative rates should be equal across a sensitive attribute (*equalized odds*).

See Figure 3 for visual explanations of statistical parity and equalized odds. We used similar visuals in our interviews, as well.

Second, we ask stakeholders to make pairwise comparisons again with cases *showing* the algorithmic prediction (Figure 2b). Participants compare cases that are selected randomly from the dataset. We ask users to identify and explain (pairs of) cases that are being treated unfairly. We also asked them to evaluate if the algorithm predictions are in general biased according to their fairness notions.

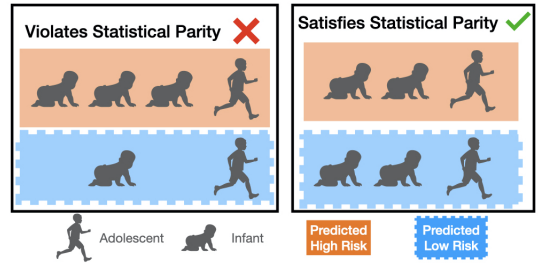

(a) Two examples which violate and satisfy statistical parity between infants and adolescents (two subgroups along the sensitive attribute *victim age*), respectively. The orange box with no fringe on top contains children predicted to be at high risk of maltreatment; the blue fringe box at bottom is low risk prediction. On the left, the proportion of high risk predictions for infants is 75%, whereas for adolescents this rate is 50%. On the right, the proportions of high risk predictions for infants and adolescents are 50%.

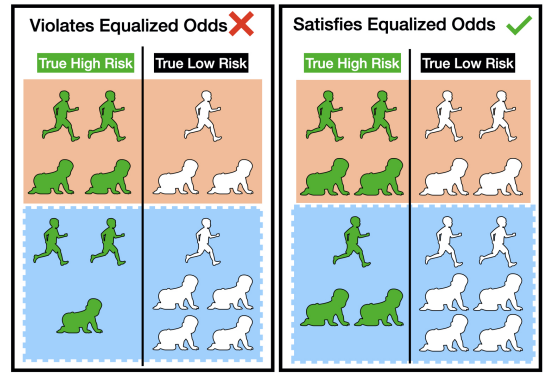

(b) Two examples which violate and satisfy equalized odds between infants and adolescents, respectively. Green children are truly at high risk of maltreatment; white are truly at low risk. False positives (children who are truly at low risk of maltreatment are predicted to be at high risk) are in the upper righthand corner; false negatives (children who are truly high risk are predicted to be at low risk) are in the lower left corner. In the left example, the false positive and false negative rates for infants are 50% and 20%, respectively; for adolescents these rates are 33.3% and 66.6%, respectively. On the right, for both infants and adolescents, the false positive and false negative rates are 50% and 66.6%, respectively.

Figure 3: High risk prediction rates must be equal along a sensitive attribute to satisfy statistical parity. False positive and false negative rates must equal to satisfy equalized odds.

Third, we ask stakeholders to use the similarity comparison view to compare reference cases with all the other cases in the data (Figure 2c). We ask stakeholders to define their own similarity metrics by ranking the importance of each feature in determining similar pairs. We then ask participants to identify cases that should be prioritized by the algorithm. We also invite participants to identify pairs which are similar to each other, but received different predictions by the algorithm. Stakeholders are free to explain the reasons behind their selections, and the information they rely on to identify them.

---
[5]Though there may be significant problems with unawareness as an approach to fairness (as noted in Section 2), we think it important to gather stakeholder beliefs about it, as it is a common and widely-used policy (e.g. color-blind policies).

Lastly, we show stakeholders the group view of the interface (Figure 2a). Stakeholders can define the groups they want to inspect, and see the algorithm's performance on the groups. We ask stakeholders to explore the groups and subgroups they are most concerned with. If participants believe any particular groups (and subgroups) are being treated unfairly, we ask follow up questions to probe the reasons for this belief.

Throughout the interview, participants are encouraged to share their views on the cases before them even if those views do not reflect perceptions of fairness per se. Participants may indicate, for instance, that they are uncomfortable with the use of algorithms in certain cases, that particular case characteristics are of paramount importance to the decision-making process, or that having model explanations would improve their understanding of the tool. This is all valuable, actionable feedback that may be incorporated into the algorithm re-training process.

## 4 USER STUDY: ELICITING SUBJECTIVE FAIRNESS NOTIONS IN CHILD MALTREATMENT PREDICTION

### 4.1 Background: Predictive Tools in Child Welfare and Related Contexts

While the use of predictive models in critical societal domains has only recently begun to receive widespread attention from the computer science community, predictive "risk assessment tools" have a long history in child welfare and beyond. Machine learning tools of the kind we discuss in this paper fall into a family of methods traditionally referred to as 'actuarial risk assessment.' The term 'actuarial' is used to indicate that a tool relies on associations inferred from data between an outcome and so-called risk factors (i.e., input features). This terminology is used to contrast with, 'clinical risk assessment', also known as professional judgment, in which experts subjectively assess risk. One of the earliest actuarial risk assessment tools was developed by Burgess [12] to calculate the recidivism risk for offenders being released from Illinois state prisons. Actuarial risk assessment instruments are now widely used throughout the criminal justice system, from pre-trial [20, 21], to sentencing [43, 55] to probation and parole [7]. They are also used in academic advising, healthcare, welfare allocation, homelessness services, and many other settings [6, 14, 44, 67].

Over the past couple of decades, many child welfare agencies have incorporated actuarial risk assessment tools—or hybrid models that combine prediction with professional judgment—into various stages of the child protection decision-making process [35, 69]. While the most widely-used tools take the form of simple point systems that consider only a handful of manually-entered factors, machine learning models such as neural networks have been considered since at least the early 2000's [54]. Contemporary tools such as the AFST differ from the majority of existing tools in that they rely on a much larger set of features that are automatically populated from multi-system administrative data. This obviates the problem of inter-rater reliability, wherein different users may have different assessments of manually-entered features in a manner that results in different risk scores. But it leaves open the possibility of more systematic errors potentially going undetected for long periods of time [18]. See Figure 4 for further explanation of the child welfare screening process used at Allegheny County Department of Human Services (DHS).

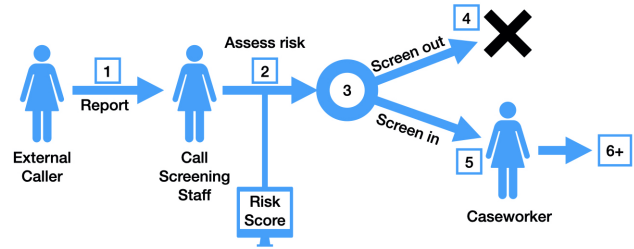

**Figure 4: Allegheny DHS screening process.**[6] **Step 1: The external caller, e.g. the child's teacher, family member, calls a child welfare hotline to make a report, which directs to either a state or county hotline. The state hotline takes down the caller's information and forwards it to the county staff in step 2 for screening. The county hotline goes directly to the call screening staff (call screeners and supervisors) in step 2, where they gather information from the caller, retrieve existing information from the department's database about the case, and assess the risk of harm to the child. Here, the AFST risk score is considered. Step 3: the call screening staff determines whether to *screen out* the report —meaning that the call is not investigated further by the department (step 4)— or to *screen in*, which can entail assigning a caseworker to the case and investigating further (step 5). Steps 6+: the case may be referred to a caseworker and/or other child welfare staff (supervisors, administrators, etc.), where they might proceed in a number of directions, e.g. further observation, investigation, or intervention (e.g. removal of the child or referral to other governmental services).**

### 4.2 Ethical Considerations Surrounding the Use of Algorithmic Tools in Child Welfare Decision-Making

The use of algorithmic decision support tools in the child welfare context is a contentious issue. For instance, there is the possibility that communities of color and families experiencing poverty may be disadvantaged by virtue of having more comprehensive data available on them in the government administrative data systems used to evaluate algorithmic risk scores. Such concerns have been given voice by authors such as Virginia Eubanks, who in her book *Automating Inequality* [25] argues that such tools oversample the poor and present a fundamentally flawed approach to improving child welfare decision-making. Similar objections have been raised by Richard Wexler, a long-time critic of algorithmic tools in child welfare.

---
[6]This visual was constructed according to resources on AFST and the screening process at DHS' website at www.alleghenycounty.us/Human-Services/Programs-Services/Children-Families/When-a-Report-is-Made.aspx and www.alleghenycounty.us/Human-Services/News-Events/Accomplishments/Allegheny-Family-Screening-Tool.aspx.

While child welfare agencies are sensitive to these concerns,[7] many also believe that data-driven decision-making approaches hold the promise of significantly improving decisions and family outcomes. Few would argue against using all available resources to promote child safety. Indeed, it can be viewed as unethical to knowingly do otherwise. Administrative system data is one increasingly available resource, but it is one that is challenging for human decision-makers to make effective and systematic use of in every instance. This is where algorithmic tools enter.

In exploring how to realize the potential upside of such tools, it is essential that agencies take a rigorous approach to development, deployment, and evaluation, and that this approach be informed by ethical considerations. Allegheny County's work on the Allegheny Family Screening Tool (AFST) used for call screening, for instance, involved both a pre-deployment ethical analysis conducted by researchers Tim Dare and Eileen Gambrill [69] and an independent post-deployment impact evaluation [27]. Studies of affected community perspectives [11] have also found that certain proposed uses of algorithmic tools are viewed by families and child welfare workers as providing considerable benefit. However, study participants voiced concern over the potential for such tools to exhibit biases and emphasized the need for a human-in-the-loop approach.

Our proposed elicitation framework is intended to respond to the clear need for algorithmic systems in sensitive domains to reflect relevant fairness and equity desiderata. We hope that the methodology we propose can work in concert with participatory design, community engagement, and impact evaluation strategies to develop tools that achieve meaningful legitimacy and demonstrably improve child and family outcomes.

### 4.3 Context and Data

We evaluated our framework in a real-world, high-stakes context—child maltreatment prediction. Our study was motivated by recent efforts by child welfare agencies around the country to incorporate algorithmic decision support tools into their existing processes. Our study was most closely related to AFST, which since August 2016 has been used during child abuse call screening in Allegheny County, Pennsylvania [69]. The AFST score is based on data related to the victim child(ren), parents, legal guardians, perpetrators, prior child welfare history, criminal history, and use of public assistance. Call screeners are presented with an AFST score for each referral. Due to the sensitive nature of the data, our study relied exclusively on synthetic data based on the real dataset provided by the Allegheny County Department of Human Services. We also converted the AFST score into binary labels—*high risk* and *low risk* cases.

### 4.4 Study Design

We recruited two groups of stakeholders for the user study: (1) social workers with experience of investigating allegations of child abuse and (2) parents. To recruit the social workers, we reached out to the departments of social work of four public universities in the US, which helped us send out the recruitment emails to their undergraduate and graduate students. To recruit the parents, we posted recruitment messages on the social media of the authors and sent out recruitment emails to parent groups. We recruited 12 participants in total for the study, and Table 1 shows the details of the participants.

---
[7]See, e.g., Allegheny County's response to Eubanks: https://www.alleghenycounty.us/WorkArea/linkit.aspx?\LinkIdentifier=id&ItemID=6442461672

| ID | Age | Gender | Race | Social worker experience | Enrolled in social work programs |
|---|---|---|---|---|---|
| S1 | 25-34 | Woman | Latinx | Yes | Yes |
| S2 | 25-34 | Woman | Black | Yes | Yes |
| S3 | 25-34 | Woman | White | No | Yes |
| S4 | 18-24 | Non-Binary | Other | No | Yes |
| S5 | 18-24 | Woman | White | No | Yes |
| S6 | 18-24 | Woman | White | Yes | Yes |
| S7 | 25-34 | Woman | White | Yes | Yes |
| S8 | 18-24 | Woman | White | Yes | Yes |
| ID | Age | Gender | Race | No. of children | Children age(s) |
| P1 | 25 - 34 | Man | Asian | 2 | 2 or younger, 3-7 |
| P2 | 45 - 54 | Woman | Asian | 2 | 13-17, 18+ |
| P3 | 45 - 54 | Woman | Other | 1 | 18+ |
| P4 | 45 - 54 | Woman | White | 3 | 18+ |

Table 1: Participant Summary

We conducted the user studies over video chat. In each study, the researcher first gave an introduction of the study and an overview of how to use the interactive interface described in Section 3.2. Then, we invited each participant to use the tool to explore the child maltreatment prediction data. We followed the interview protocol introduced in Section 3.3 to elicit participants' fairness notions surrounding the algorithm's decisions. Participants shared their screen in the video chat so that researchers could see the same information in the process. Participants were also encouraged to think aloud during the study. Each user study lasted for about 90 minutes. Each participant was compensated with a $30 Amazon gift card. The user study was reviewed and approved by the Carnegie Mellon University Institutional Review Board.

### 4.5 Data Analysis

All study sessions were audio-recorded with consent from the participants. The first two authors transcribed all 12 interviews with 20.5 hours of recorded audio. We employed a qualitative, grounded theory analysis to inductively analyze our data and generate the findings and insights from the interviews. We adopted Charmaz's approach of grounded theory analysis which allows us to consider prior ideas and theory in the analysis [15]. We open coded interview transcripts, held team meetings to discuss emerging themes and ideas, and iterate on our codebook. We describe the findings of our analysis in the next section.

## 5 RESULTS

In this section, we summarize participants' viewpoints on both group and individual fairness gathered throughout the interview process. Overall we see no difference in patterns of responses between social workers and parents. Throughout this section, we follow the prior literature and refer to the three group fairness approaches by these abbreviated phrases: *unawareness* means to leave a sensitive attribute (e.g. race) out of the model; *statistical* *parity* means to equalize the positive classification rates between groups within a sensitive attribute (e.g. different racial groups); and *equalized odds* means to equalize the false positive and false negative rates between groups within a sensitive attribute. The highlights of our results are as follows:

(1) Among the three group fairness approaches, equalized odds was the most supported group fairness criteria (66.7%), followed by statistical parity (43.3%) and unawareness (41.7%).
(2) Even though equalized odds was the most supported, there were nuances within this, e.g. many participants were willing to accept disparities in accuracy across groups rather than sacrifice overall accuracy.
(3) Even though there are heated discussions around the over-representation of Black children in the child welfare system [22], participants thought that statistical parity was not necessarily fair, though it is a good goal.
(4) Participants thought awareness could both address or enforce systemic discrimination.
(5) Among the individual fairness comparisons, we did not observe unanimous agreement between participants most of the pairs.
(6) Participants maintained consistent responses to each group fairness question across different protected attributes (i.e. victim age, victim gender, family race, use of public assistance, and perpetrator gender).
(7) Participants interpreted our fairness questions differently and experienced cognitive overload when examining cases with a high number of different attributes, leading to additional challenges.

## 5.1 Group Fairness Approaches

*5.1.1 Summary of Group Fairness Choices.* We asked the participants whether three common group fairness approaches (unawareness, statistical parity, and equalized odds) are appropriate for each of the five sensitive attributes (victim age, victim gender, family race, use of public assistance service and perpetrator gender). Particularly, for a given sensitive attribute, we asked whether the following criteria should be met for the algorithm to be fair:

(1) the sensitive attribute should not be considered for decision-making (*unawareness*);
(2) the high-risk predictions be the same rate across the sensitive attribute (*statistical parity*); or
(3) the predictions should be equally accurate at identifying low- and high-risk cases across the sensitive attribute (*equalized odds*).

See Figure 3 on page 5 for visual explanations of statistical parity and equalized odds.

Among the three group fairness criteria, equalized odds was the most preferred group fairness criteria (66.7%), followed by statistical parity (43.3%) and unawareness (41.7%) (see Figure 6).

---

[8] The observant reader might notice that the equalized odds responses are split 8-4, which is the same as the proportion of social workers versus parents. This is simply a coincidence. The four participants who responded 'No' were half social workers and half parents. In general, we tested for clustering of our social workers' and parents' responses to all questions to see if there were any divisions between the two groups of participants and found little.

(a) Responses to the question *"Should a fair algorithm be aware of a given sensitive attribute?"* (awareness)[a]

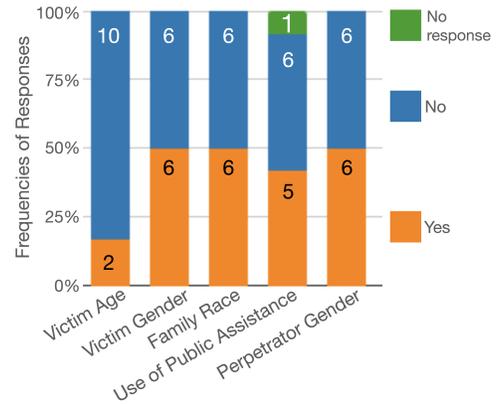

[a] 'No' means the participant believes the algorithm should be *unaware* of the attribute. 'Yes' means the participant believes the algorithm should be *aware* of it.

(b) Responses to *"Should a fair algorithm classify equal proportions of cases as high risk between subgroups within a given sensitive attribute?"* (statistical parity)

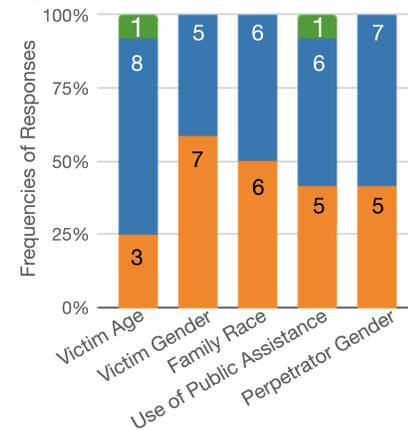

(c) Responses to *"Should a fair algorithm have equal accuracy (false positive and false negative rates) between subgroups within a given sensitive attribute?"* (equalized odds)[8]

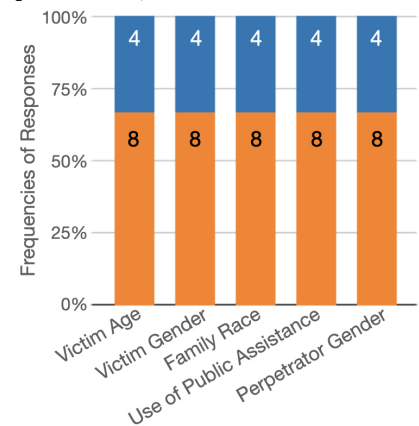

Figure 5: Frequencies of responses to the three group fairness questions for sensitive attributes victim age through perpetrator gender. Also, e.g. infants and adolescents are two subgroups within the victim age sensitive attribute.

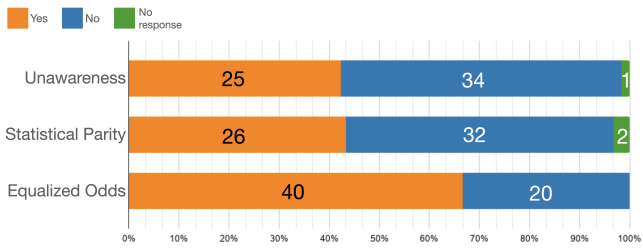

**Figure 6: Frequencies of responses to the three group fairness questions. See Figure 5 for further explanation of questions asked.**

*5.1.2 Reasons for & against Unawareness.* Our participants think unawareness is an appropriate fairness criteria in 41.7% of situations. 6 out of 12 participants believed that algorithm should be unaware of victim gender, family race and perpetrator gender; only 2 out of 12 participants believed that the algorithm should be unaware of victim age. (see Figure 5a).

**Algorithms should be aware of important predictive attributes.** Many participants disfavored unawareness, i.e. endorsed awareness, for sensitive attributes they thought were important indicators of risk. For example, S2 thought age should be taken into account, since *"when you're a younger child that doesn't have language, you definitely are higher risk than a child that does have language"* (S2). Many other participants reiterated that victim age is an indicator of language and, thus, of risk. This is one reason why we saw only 2 of 12 participants endorsing unawareness for victim age (see Figure 5a). Others thought an algorithm should be aware of *only* important predictive attributes. For example, S7 –who endorsed unawareness for victim age– said, *"I wouldn't want to like teach the algorithm to prioritize [a case] based on age... I would want it more about the type of alleged abuse,"* indicating that type of abuse is an important predictor, whereas victim age is not.

**Awareness could reinforce systemic discrimination.** Some participants endorsed unawareness, because they were concerned that awareness of sensitive attributes would lead to systemic discrimination. For example, P3 thought that the algorithm ought to be unaware of family race, since awareness *"opens up... room for systemic racism"* and would *"bring unconscious biases"* (P3). Furthermore, some suggested awareness of sensitive attributes to audit the algorithm's decision, but not when making predictions. P3 said that the algorithm should be aware of family race *"only to report certain disparities"* between races (P3). Similarly, S4 endorsed unawareness so that the algorithm would not inherit biases based on gender, saying: *"There is a big bias on gender. Everybody has different opinions on what gender is [and] what entails gender"* and a fair algorithm should be unaware of gender *"to erase all of that"* (S4). One participant, S7, endorsed unawareness to prevent bias, not just for marginalized groups, saying that an algorithm that is aware of victim gender *"would definitely learn to prioritize female victims"* over male victims, so it should be unaware of victim gender (S7).

**Awareness could help address systemic discrimination.** Other participants the opposite: that an algorithm should be *aware* of sensitive attributes, because awareness could help address and correct for historical disparities. For example, S1 thought an algorithm should be aware of family race in order to *"negate any... racial biases"* (S1). S4 thought that family race should be one of the most important predictive factors, saying *"people of color are disproportionately affected and they always have been;"* so *"if you're a person of color, you should be prioritized over white people,"* because fair predictive systems *"should be targeted to help people who we... know need the help"* (S4).

**Endorsing awareness may come with apprehension.** Some participants identified a reason for unawareness, but ultimately decided against it. For example, S6 thought that the algorithm ought to be aware of all sensitive attributes, saying, *"The more information you have, the better. But I can understand how that can also lead to bias"* (S6). S5 said that victim *"age is an important factor,"* but it should not be overemphasized as a predictor: there is *"room for algorithm error if, it's focusing so much on age, rather than [on] other factors"* (S5).

*5.1.3 Reasons for & against Statistical Parity.* Our participants thought statistical parity was a slightly more appropriate fairness approach than unawareness, supported in 43.3% of the cases. Similarly, statistical parity was most supported for victim gender (7/12) and family race (6/12) and least supported for victim age (3/12) (see Figure 5b).

**Statistical parity is a good goal, but is not necessarily fair.** While many participants wanted to see similar positive predictive rates across these groups, they also thought that disparities in the prediction rates do not indicate that an algorithm is unfair. For example, with regards to high-risk prediction rates, S5 said, *"You'd want to see similar numbers achieved, but if the numbers aren't quite the same, I think that's also okay as long as you're still detecting fairly"* (S5). P1 said, *"I wouldn't say that if the algorithm predicts differently across different races, then it's an unfair algorithm"* (P1).

**Statistical parity overlooks contextual differences between cases.** Some participants who disfavored statistical parity reasoned that equalizing positive classification rates may overlook contextual differences specific to each case. For example, P2 said that statistical parity *"is arbitrary... It depends on what the cases are"* (P2).

**Statistical parity overlooks different base rates between groups.** Many participants recognized that different groups may have different base rates of being at high risk of abuse, especially among different victim age groups. As a result, many expressed apprehension towards statistical parity. For example, P4 and S4 disfavored statistical parity among victim of different ages, because, as P4 said, *"I think the prediction [rates] will be higher in the lower age group, so [different age groups] shouldn't be looked at the same"* (P4). S8 explained why they disfavored statistical parity using a scenario where the base rates of high-risk classification among different ages are different due to circumstances unrelated to abuse: *"Children under five are not school age yet. So, there may be fewer adult eyes on the child. There might be more eyes on older children, so there might be higher rates of referrals; but, that doesn't mean there's higher rates of abuse"* (S8). P1 disfavored statistical parity between victim genders, because *"there's probably difference, in terms of the risk, between the two genders"* (P1). When asked whether statistical parity is fair, S7 said, *"Yes, with an asterisk, because I do think in cases of alleged sexual abuse, I would expect to see a higher percentage of high risk for female victims"* (S7).

Furthermore, some participants advocated for something like calibrating the rates of high-risk predictions to the base rates. For example, S3 said, an algorithm *"should be making high-risk predictions based on what demographic data says is the most at risk"* (S3). They even recognized that *"there might be certain populations that are under-reported are over-represented, but I think that to the best of our ability high risk predictions from the algorithm should match... demographic information"* (S3). This comment is particularly interesting, considering concerns over the child welfare system stemming from over-representation of Black children in foster care [22], as will be further explained in Section 6.1.1.

**Mandating statistical parity is not fair.** Some participants disfavored statistical parity, because they did not think it was fair to mandate this condition be met. For example, if the algorithm were focused on achieving statistical parity, S8 said they *"would worry that [the algorithm] would be focused on meeting a number"* (S8). P2 said, *"The algorithm should not try to balance [the rates of classification] out so that it appears to be fair... Trying to balance out the rates of high-risk predictions means... you're manipulating the situation"* (P2).

*5.1.4 Reasons for & against Equalized Odds.* Equalized odds was the most supported fairness approach: participants thought it was an appropriate fairness approach 66.7% of the time. There were no differences in the frequencies of support across all sensitive attributes (see Figure 5c).

**Equalized odds aligns with existing fairness beliefs.** We found that participants tended to agree with equalized odds since it is more closely aligned with their existing fairness beliefs—many participants expressed that they want the accuracy of the algorithm to be as high and as even as possible. For example, S4 said, *"I think the goal is for [the algorithm] to be 100% accurate. But if I had to choose one over the other. Yeah, I would want [the accuracies between groups] to be [equal]. It should have the same accuracy and it should be high"* (S4). Numerous participants echoed almost this exact sentiment, e.g. S6 said, *"I would want for it to have the same accuracy and for the accuracy to be really high across all groups"* (S6). This is consistent with equalizing the true positive and true negative rates across groups, which equalized odds entails.

**Accuracy should not be sacrificed to achieve equalized odds.** However, an algorithm cannot always achieve equalized odds. Participants also discussed trade-offs they are willing to accept if equalized odds is not attainable. To achieve equalized odds, one can lower the accuracy –i.e., increase the false positive rate or false negative rate– for the group where the algorithm originally performs better. The participants generally disagree with such practices to attain equalized odds. S8 said, *"I wouldn't want to lower the accuracy of a group. That seems counter-intuitive. You want it to be accurate as much as it can"* (S8). For this reason, S8 disfavored equalized odds. Other participants, however, thought that the accuracies should not be lowered and endorsed equalized odds. For example, S5 endorsed equalized odds for all sensitive attributes, but responded, *"I don't think the accuracy should be lowered"* (S5).

It's worth noting that other participants, such as P3, were fine with lowering the accuracy of one group to match the other in order to achieve equalized odds.

**Improvements in accuracy should help as many high-risk people as possible.** In the case that equal odds cannot be achieved, some participants wanted the algorithm to improve accuracy for larger groups of people in order to help as many people as possible. For example, S6 said, *"I obviously want [the algorithm] to be as accurate as possible across all groups. I would want it to have the same accuracy and for the accuracy to be really high across all groups. But, when there's different accuracies, you would want a higher accuracy for groups that are bigger. Hopefully, you would help the most... children as possible."* (S6). S6 clarified that not only large groups, but those who are high-risk should be prioritized: *"If the most abused or neglected group is like Hispanic girls you would want that [group] to have higher accuracy than like white boys"* (S6). S4 also expressed similar ideas, saying *"I think [the algorithm] should be more accurate on non-white races and ethnicities than white ones, just because I feel like they are people of color and are more at risk"* (S4).

**Equalized odds hides historical inaccuracies.** Instead of equalized odds, some participants would rather be aware of an algorithm's historical false positive and false negative rates for a given group. As a social worker, S8 thought that mandating equalized odds hides where the algorithm is more or less accurate: instead, they said, *"I would want to know where [the algorithm] was less accurate, so that we could be looking at what's getting in the way. And how can we improve"* (S8). Participants recognized the role of the human decision-maker that considers the algorithm's prediction as a non-binding recommendation. For example, P1 did not think that you have to make *"your decision purely based on the algorithm: you can know that the algorithm has different prediction accuracy across different factors"* (P1).

It's worth noting that other participants recognized the importance of knowing historical inaccuracies, yet endorsed equalized odds. For example, when asked whether differences in false positive or false negative rates across groups were unfair, S5 (who endorsed equalized odds for all sensitive attributes) responded, *"it depends on why the difference is there"* (S5).

*5.1.5 Group fairness guided by data and research.* In addition to the established group fairness definitions that often ask for parity between the metrics, participants also discussed an additional guideline that the algorithm should be following. S3 expressed the idea that, instead of having statistical parity between the groups, the positive prediction rate should be matched to the historical demographic data: *"I think that [the algorithm] should be making high risk predictions based on what demographic data says is the most high risk. The high risk predictions... from the algorithm should match as close as possible to the demographic information that showed you who was being victims of abuse"* (S3). This approach, however, has the risk of potentially reinforcing historical injustice. To put that into practice, S3 thought that statistical parity was not appropriate across perpetrator genders: *"I think it should match demographic information and demographic information has shown us that perpetrators are more likely to be male"* (S3). When asked which sensitive attributes equalized odds is an appropriate fairness measure for, S2 said, *"my thought process is that I want to make sure that whatever decision that I would make is informed by research"* (S2).

At the same time, when discussing if the algorithm should consider a feature in its prediction, participants thought that if the algorithm's decisions are supported by research findings and existing data, then this is a signal that the algorithm is fair. For example, when asked about which sensitive attributes a fair algorithm should be aware of, S7 said that *"any factor that's supported by research as being linked to a higher frequency of outcomes should be considered by the algorithm... If it's not supported in the research, then I don't think it needs to be introduced"* (S7).

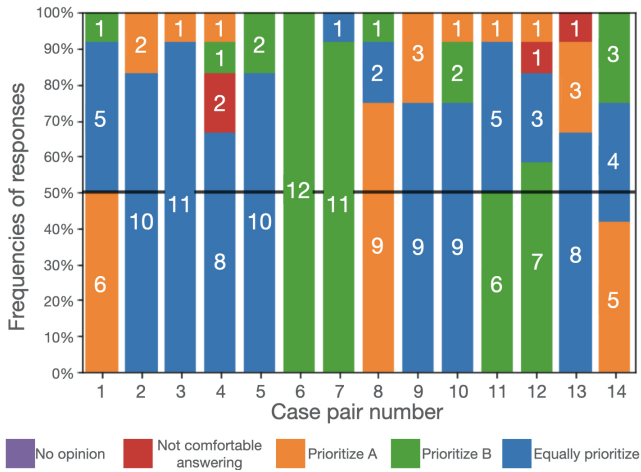

Figure 7: Frequencies of responses by case number, where the case pair numbers correspond to differences between the pair as follows. Case 1: Victim age; 2: Victim gender; 3: Family race; 4: Use of public assistance; 5: Perpetrator gender; 6: Allegation type, Perpetrator age; 7: Family race, Referral history; 8: Use of public assistance, Victim age, Reporter type; 9: Victim age, Perpetrator age (Perpetrator not related to victim); 10: Victim age, Perpetrator age (Perpetrator related to victim); 11: Number of parents, Region wealth, Perpetrator relationship to victim; 12: Region wealth, Use of public assistance, Referral history; 13: Family race, Region wealth, Use of public assistance; 14: Number of parents, Victim age, Victim gender

## 5.2 Individual Fairness Notions

*5.2.1 Summary of Individual Fairness Choices.* Participants use the case-by-case view (Figure 2b) to evaluate different pairs of cases treated by the algorithm. For the first 14 pairs of cases, the actual algorithm predictions are not shown and participants discussed their opinions on how the algorithm should be treating these cases. Figure 7 shows participants' responses to the 14 pairs of fixed cases. For each pair of cases A and B, we offered participants the following five options to choose from:

(1) *(Equally prioritize)*: cases A and B should be given the same classification (either high- or low-risk);
(2) *(Prioritize A)*: only case A should be classified as high-risk;
(3) *(Prioritize B)*: only case B should be classified as high-risk;
(4) *(Not comfortable answering)*; or
(5) *(No opinion)*.

**We did not observe unanimous agreement.** Though there is frequent majority consensus among the individual fairness comparisons —only three case pairs (1, 11, and 14) do not exceed a majority (above the 50% line),— we only saw unanimous agreement between participants in one (case pair 6) of fourteen cases. This is significant, because if we expected there to be one best fair response for each case, then (even among our small number of participants) we might have expected unanimous consensus among more case pairs. Yet, only five pairs had 10 or more participants who responded the same.[9] One common theme between these 5 pairs was that the only differences between the pairs were allegation type, gender, race or a combination of them. Most agreed on *equally prioritizing* those in case pairs 2, 3, and 5, wherein the cases differ only along victim gender, family race, and perpetrator gender, respectively. In case pair 6, the pairs differed only along allegation type and perpetrator age: all participants voted to prioritize the case with the more serious allegation. In case pair 7, the pairs differed only along referral history and family race: all participants but one voted to prioritize the case with with more prior referrals.

In all the other cases, responses were more contentious. In case pair 1, where the cases differ only in victim age, we saw an almost even split between prioritizing the case with younger age (6 participants) and prioritizing both cases equally (5 participants). For case pair 11, we saw another even split between prioritizing both cases equally (5 participants) and prioritizing the case with the single parent, non-parent perpetrator in a less wealthy region (6 participants). For case 14, we saw an even split between prioritizing both cases equally (4 participants), prioritizing the case with a younger, male child and two parents (5 participants), and prioritizing the case with an older, female child and single parents (3 participants).

*5.2.2 Examples of agreed pairs vs contentious pairs.* For the highly agreed-upon pairs, all participants had a strong and clearly-defined consensus on how these features affect the prioritization that a maltreatment case should receive. In one situation, the implication of a single feature was so clear that the participants universally agree on prioritizing on the case with the more severe allegation type, and do not need to look at the other less significant features: *"I think for me immediately what gets me to prioritize B over A is the allegation type. For me I think allegation types should be prioritized over the age. So that's the first thing... I would screen... and I don't really think I have to look at the age anymore"* (S4).

In some scenarios, the decision was a lot more contentious, especially when participants disagreed on the implications of a specific factor. For example, participants disagreed about whether differences in victim's age was a strong enough reason for an algorithm to prioritize two cases differently. Some participants thought age difference alone should not be a reason to differentiate two cases: *"Now the only thing that's different is the age, so... it doesn't really feel right to say one should be prioritized over the other, necessarily... I don't really think [age] should make a difference"* (S5).

Some participants thought younger children were generally more vulnerable than older children and, therefore, should be prioritized first by the algorithm: *"the twelve-year-old is going to be able to say what's happening easier than the four year old, because the twelve-year-old girl would have a better idea what's wrong and right than a four year old."* (S6).

---
[9]The threshold for unanimous consensus is not entirely clear. We point out this threshold of 10 participants in agreement as an example. However, if we chose 9 participants as a threshold, the number of case pairs increases to eight out of fourteen. The point still remains that full unanimous consensus was not reached in most cases.

*5.2.3 Heuristics on how individual cases are compared.* When navigating through the case-by-case and similarity comparison view, participants were asked to make decisions for child maltreatment cases based on the limited information available to the algorithm. As a result, some information that the participants wanted to know might not have been available when making the decisions. We observed two different types of heuristics when the participants were comparing different cases side-by-side.

**Participants constructed stories upon available case information.** Some participants used the information available in each case to reconstruct the story the victim is facing in each case. They would attempt to compare and visualize which of the cases are facing a more imminent risk. When there was information that was missing, the participants would come up narratives of the potential scenarios. For example, S1 said, *"A really important factor definitely would be like what kind of abuse, we're talking about here like is it a neglect case? Is it a case of a babysitter not paying attention to the kids or is this like, if it was something physical or sexual then it's obviously one I would be much more concerned about"* (S1). These participants weighed the risks and probabilities of all these alternatives in order to make a final decision.

**Participants focused on the differences between cases.** The other way participants compared the cases was to focus primarily on the differences in the features between the two cases. For these participants, they would first look through the features one by one and identified the first differences they spotted between the cases. Based on the differences, the participants would weigh in how each of the differences affected the differences in their potential risk, then made a final decision based on that. For example, S6 explained how they chose to prioritize a cases based on the difference in the number of prior referrals: *"the only difference really is the income and the fact that this one had a referral. I guess I would choose the one that's already had a referral"* (S6). Participants prioritized cases with features that they thought were more significant.

We also asked participants to rank the importance of each feature in determining if two cases are similar or not. We found that participants were more likely to use the features they ranked as more important as the primary factor in deciding which case should be prioritized by the algorithm.

When facing situations where the differences were of the same significance to them, these participants would simply count the pros and cons for prioritizing each case. For example, S4 reasoned, *"So going off the race. I would prioritize the right [i.e. case B]. But, then because of public assessor, [I would] prioritize the left [i.e. case A]. And then on the socioeconomic factors I would prioritize have left. So, because it's two to one, I would prioritize case A over case B in this situation"* (S4).

## 5.3 Internal Consistency in Participants' Responses

*5.3.1 Internal consistency in group fairness questions.* When asked about group fairness questions, each participant often answered similarly across all sensitive attributes. For each group fairness question when participants responded to all five sensitive attributes (victim age, victim gender, perpetrator gender, family race, and use of public assistance), they responded to a mean of about 4.4 out of 5 questions (87.8%) the same. For equalized odds, all participants answered only either 'yes' or 'no' over all sensitive attributes.

This indicates that our participants had internally consistent beliefs across protected attributes. Participant responses indicate that this was because participants had a common reason for favoring or disfavoring a group fairness approach across all protected attributes. For example, when asked whether equalized odds across different ages was fair, S8 gave a reason for responding 'no'; then, they said they *"feel similarly regardless of the identity"* (S8). This kind of response was common among participants: they held the same reason, so they answered the same across all protected attributes.

*5.3.2 Internal consistency between group fairness and individual fairness.* Many participants maintained consistency in reasoning over protected attributes between individual and group-level responses. For example, when asked about case pair 1 –which differed only based on age,– S3 thought that the younger child ought to be prioritized. Later, when asked whether statistical parity across different ages is fair, S3 answered 'no,' reasoning: *"because the cases that I looked at, I selected the lower-age children [to be prioritized]"* (S3). Participants reasoned about individual cases when asked group-level questions and vice versa. For example, S7 also reasoned about a previous pair of cases when asked whether a fair algorithm should be aware of victim gender. From the other direction, when presented with case pair 3, wherein the only difference was that one family was Caucasian and the other was African-American, S2 reasoned that *"race... holds a lot of social meaning in America"* (S2) and chose to equally prioritize these pairs as a proxy for making sure that these different races are being treated equally.

These results indicate internal consistency in reasoning across individual and group fairness. *We speculate that this is because participants have preconceived notions of fairness that run through all their answers.*

## 5.4 Challenges of Interpreting the Responses

*5.4.1 Different understanding of group fairness questions.* When participants were asked about group fairness criteria (statistical parity or equalized odds in particular), they gave reasons that indicated different understandings of these questions. As a result, the meanings of 'yes' and 'no' answers to these questions were varied.

**Some understood statistical parity and equalized odds questions as asking about sufficient conditions.** Some participants understood these questions as asking, *Are statistical parity or equalized odds sufficient conditions for fairness?* For example, when asked whether statistical parity was fair, S3 said they *"want to see similar numbers* [i.e. rates of positive classification] *achieved, but if the numbers aren't quite the same,... that's also okay"* (S3). Thus, S3 thinks that statistical parity is a sufficient condition for fairness, because if the algorithm meets statistical parity, then this is a fair outcome. Based on their reasoning, if they understood the question as asking whether statistical parity were a *necessary* condition, they may have answered 'no'.

**Some understood statistical parity and equalized odds questions as asking about necessary conditions.** Some participants understood these questions as asking, *Are statistical parity or equalized odds necessary conditions for fairness?* For example, when asked whether a fair algorithm should achieve statistical parity, P1 said, *"The question [is] weird because if I say 'No,' what I'm saying is that a fair algorithm shouldn't make high risk prediction at the same rate across groups"* (P1). P1 answered 'no' to all statistical parity questions, because, even though they indicated that *ideally* the algorithm should classify different groups at similar rates, it *"does not have to"* to be fair.

**Some understood statistical parity and equalized odds questions as asking whether they should be mandatory.** Another group of participants understood these group-fairness question as asking, *Should the algorithm be mandated to fulfill statistical parity or equalized odds in order to be fair?* For example, P2 answered 'no' to all statistical parity questions, because *"the algorithm should not try to balance* [the rates of classification] *out so that it appears to be fair... Trying to balance out the rates of high risk predictions means... you're manipulating the situation"* (P2). P1 and S8 answered 'no' to all statistical parity and equalized odds questions: S8 said they *"would worry that [the algorithm] would be focused on meeting a number"* (S8); P1 said that *"deliberately changing the algorithm to be lower accuracy so that [the classification accuracies across protected groups] match... just doesn't make sense"* (P1). These participants thought that if group fairness constraints were met by mandating the algorithm to do so, these were not necessarily fair situations. Thus, their 'yes' or 'no' responses have no bearing on whether statistical parity or equalized odds would be fair if they were met *without being mandated*. It's possible they would have answered 'yes' if we asked this.

These varied understandings of group fairness questions illustrate that group fairness approaches can be complex and ambiguous. Additionally, questions about group fairness definitions should specify whether they are about necessary and sufficient conditions, as well as whether the question is asking to mandate these rules or not.

#### 5.4.2 Unfairness versus wrong prediction.
When participants were making pairwise comparisons between the cases, we found that some participants perceived the algorithm as making an unfair decision if they thought the algorithm made a mistake for a single case, rather than a pair of cases. For example, S6 explained why she thought a particular case (currently predicted as low-risk) was treated unfairly by the algorithm: *"I would say, like this one is unfair. They have one parent and allegation is parents, substance abuse, it should be a higher risk prediction"* (S6). When asked if the decision was reached based off of a comparison with a similar case, S6 explained *" I just think it's unfair. This kid only has one parent and the allegation is substance abuse, that would mean that parent is not really able to take care of that kid, and the kid doesn't have any other parents there"* (S6). This participant did not identify unfairness based on comparing individual case pairs, but rather by a perceived misclassification of a single case made by the algorithm.

#### 5.4.3 Cognitive overload.
The notion of individual fairness calls for comparison between individuals to determine if they should be treated similarly. However, we found that the comparison was not always easy for human stakeholders, especially in real-life scenarios where each case included numerous features. Participants explained that having to objectively compare the effect of multiple differences between individuals can be challenging. *"I think it's tricky to compare things this way,... because of the multiple factor difference. It's hard to say. If you can control all of them and only one is changing, then that might be easier"* (P1).

We did not measure the cognitive load of the participants directly; thus, we cannot tell if the task of comparing individual pairs was mentally taxing. The participants' response reflected that it would have been easier for them to compare pairs with limited differences (e.g. selecting pairs with high similarity in similarity comparison view), than to compare randomly-selected pairs (e.g. randomly pairs in case-by-case view).

## 6 DISCUSSION
## 6.1 Implications for Model Design and Development

In this section, we discuss the implications of our findings for the design, development and evaluation of future predictive systems for use in child welfare decision-making. While there is existing work on how to directly incorporate pairwise comparison feedback in to model training [38, 53], their mechanisms focus on simple aggregation rules. In comparison, our framework elicits richer perspectives, especially the stakeholders' reasoning behind their responses. These findings can potentially lead to more structural changes to the risk assessment tools. We discuss three directions on how to incorporate our findings into model design and deployment.

*6.1.1 Group fairness.* One interesting take-away from our study is that broadly accepted group-level indicators of potentially unwarranted disparity within the child welfare system are not unanimously held as indicators of algorithmic bias. For instance, one of the most commonly cited indicators of bias is the over-representation of Black children among those investigated and placed in foster care [22]. In the language of group fairness, these decisions fail to satisfy statistical parity. However, as we note in our results, participants were evenly split on whether they viewed statistical parity to be a desirable algorithmic fairness property, especially for different races (see Figure 5b). This finding indicates that common measures of bias in historical decision-making may not constitute reliable fairness metrics for the purpose of algorithm design and evaluation.

Our results also do not fully agree with prior work on human perceptions of fairness [68], which suggests that statistical parity most closely matches people's existing notion of fairness. Our work differs from [68] in terms of context and user elicitation method: [68] investigates opinions on criminal risk and skin cancer screening by crowdsourcing opinions from the public; our work investigates child welfare by engaging with relevant stakeholders. These differences in methodology may explain the differences between our results and [68]. This may also indicate that the appropriate fairness approach is both dependent on context and stakeholders, which further exemplifies the need for algorithm designers to incorporate relevant stakeholders fairness viewpoints into the design process.

While we found that for child maltreatment prediction, there was no single group fairness approach that was universally supported by all participants, equalized odds aligned with the existing fairness beliefs of most participants, as they tended to want an algorithm to be as accurate as possible and similarly accurate for as many people as possible.

*6.1.2 Individual fairness.* Our participants' individual fairness feedback can be used in the model re-training process. A starting point is to follow the same approach as in [8, 38] and formulate the pairwise comparison responses as constraints on the predictive model. For example, if a participant indicated that case A should be prioritized over case B, then the model should provide a higher risk score for case A. Our study showed that there are frequent disagreements among the participants' responses, so incorporating this individual fairness feedback requires a mechanism to resolve disagreements. Prior work [53] resolves such disagreements based on the Borda rule from voting theory. Alternatively, one can also resolve such disagreements through a deliberation process among stakeholders. Furthermore, our interface provides richer information that enables feedback beyond these pairwise constraints. For example, the similarity interface in Figure 2c elicits from each participant a similarity metric, which can potentially define individual fairness criteria beyond the sets of pairwise comparisons. Another important factor is that we choose not to present all of the attributes used in child welfare screening, since it may be excessively cognitively difficult for the participants to process over 100 attributes. For any given comparison, the response of prioritizing case A over case B can be interpreted as (1) providing a higher average risk score for all cases like case A, or (2) providing higher risk scores for typical cases like case A.

*6.1.3 Beyond fairness.* While the primary focus of our study and interface design is on fairness preference elicitation, our interview protocol enables us to learn about stakeholder perspectives along other dimensions as well. For example, our study also indicates that the child maltreatment risk model and the overall decision process can benefit from the inclusion of additional attributes. From the outset of our study, participants expressed how challenging it was to make prioritization decisions based on the limited information provided. For example, S8 noted that there can be all sorts of reasons that a family might need public assistance service and, depending on what these reasons are, they could be either a protective factor or a potential indicator that the family is struggling. Having access to this additional information may help both the model and the human decision-maker in determining the risk in each specific case.

Our participants naturally put more emphasis on certain features when evaluating whether one of the two cases should be be prioritized by the algorithm. Participants universally agreed that victim age, allegation type and prior referrals were more important when evaluating which case should be prioritized. More generally, in any given context, an automated model selection procedure is prone to produce an algorithm that doesn't rely on or prioritize many of the features that expert stakeholders believe are the most important. This is problematic because model predictions that frequently disagree with users' perceptions may be viewed as not credible by those users [73]; or, worse, *users* that frequently disagree with the model predictions may be viewed as not credible. In the present context, an algorithm that is trained to prioritize attributes that stakeholders find important may see greater uptake that one that optimizes for predictive accuracy alone.

Ultimately, a predictive model should be only one of the steps in the pipeline of child maltreatment. In the end, the final decision is made by human call screeners, who look at the model's prediction as a non-binding recommendation. It is important to look at the fairness of the decisions of the entire socio-technical system, not just the predictive model within it. We argue that it is critical that the fairness viewpoints of these stakeholders be heard and incorporated in the model. At the same time, while it may not be possible for the algorithm to be fair in every possible scenario, it is more important for the users (e.g., the call screeners in the child welfare context) to recognize the limitations of the model in order to make a final decision based on the recommendation of the model.

## 6.2 Limitations

As with any study, it is important to note the limitations of this work. Since this is a qualitative study, the insights we report only represent the fairness viewpoints of the 12 participants we interviewed. Other stakeholders may hold different opinions from the participants we interviewed. The proportion of responses in our results may not reflect exact distributions. Nevertheless, our results highlight that participants did not unanimously agree on what should be considered fair, and they indicated qualitatively different reasoning for their responses.

For ethical reasons, we did not interview parents who have interacted with or are interacting with the child welfare system in Allegheny County, PA. Therefore, the demographics of our parent sample group may not reflect the demographics of parents in the child welfare system. In addition, while children are also stakeholders of the child welfare system, we were unable to interview children directly due to ethical concerns of asking children sensitive questions and their lack of understanding to the system. Finally, we primarily interviewed social workers with casework experience. Further studies which aim to capture the beliefs of all stakeholders within a child welfare department should likely include more supervisors, administrators, and executives, as well.

## 7 CONCLUSION

In this work, we present a general framework to elicit stakeholders' subjective fairness notions regarding algorithmic systems. We evaluate our framework on a child maltreatment predictive system and conduct a user study with relevant stakeholders. The interviews provide us with a comprehensive understanding of stakeholders' perspective of algorithmic fairness. We find that equalized odds is the slightly more preferred group fairness approach for child maltreatment risk assessment, but stakeholders do not unanimously agreement about individual fairness comparisons. We relate our findings to incorporating stakeholders' feedback in the design and development of algorithmic predictive systems.

## ACKNOWLEDGMENTS
We thank our anonymous reviewers, colleagues from GroupLens Research at the University of Minnesota and the HCI Institute at Carnegie Mellon University for their feedback. This work was supported by the National Science Foundation (NSF) under Award

## REFERENCES

[1] Alekh Agarwal, Alina Beygelzimer, Miroslav Dudík, John Langford, and Hanna M. Wallach. 2018. A Reductions Approach to Fair Classification. In *Proceedings of the 35th International Conference on Machine Learning, ICML 2018, Stockholmsmässan, Stockholm, Sweden, July 10-15, 2018*. 60–69. http://proceedings.mlr.press/v80/agarwal18a.html

[2] Alekh Agarwal, Miroslav Dudík, and Zhiwei Steven Wu. 2019. Fair Regression: Quantitative Definitions and Reduction-Based Algorithms. In *Proceedings of the 36th International Conference on Machine Learning, ICML 2019, 9-15 June 2019, Long Beach, California, USA*. 120–129. http://proceedings.mlr.press/v97/agarwal19d.html

[3] Yongsu Ahn and Yu-Ru Lin. 2019. Fairsight: Visual analytics for fairness in decision making. *IEEE transactions on visualization and computer graphics* 26, 1 (2019), 1086–1095.

[4] E. P. Apfelbaum, K. Pauker, S. R. Sommers, and N. Ambady. 2010. In blind pursuit of racial equality? *Psychological Science* 21 (2010), 1587—1592. https://doi.org/10.1177/0956797610384741

[5] Edmond Awad, Sohan Dsouza, Richard Kim, Jonathan Schulz, Joseph Henrich, Azim Shariff, Jean-François Bonnefon, and Iyad Rahwan. 2018. The Moral Machine experiment. *Nature* (2018), 59–64. Issue 563. https://doi.org/10.1038/s41586-018-0637-6

[6] Caroline Balagot, Hector Lemus, Megan Hartrick, Tamera Kohler, and Suzanne P Lindsay. 2019. The homeless Coordinated Entry System: the VI-SPDAT and other predictors of establishing eligibility for services for single homeless adults. *Journal of Social Distress and the Homeless* 28, 2 (2019), 149–157.

[7] Geoffrey Barnes and Jordan M Hyatt. 2012. Classifying adult probationers by forecasting future offending. (2012).

[8] Yahav Bechavod, Christopher Jung, and Zhiwei Steven Wu. 2020. Metric-Free Individual Fairness in Online Learning. *CoRR* abs/2002.05474 (2020). arXiv:2002.05474 https://arxiv.org/abs/2002.05474

[9] Rachel KE Bellamy, Kuntal Dey, Michael Hind, Samuel C Hoffman, Stephanie Houde, Kalapriya Kannan, Pranay Lohia, Jacquelyn Martino, Sameep Mehta, A Mojsilović, et al. 2019. AI Fairness 360: An extensible toolkit for detecting and mitigating algorithmic bias. *IBM Journal of Research and Development* 63, 4/5 (2019), 4–1.

[10] Reuben Binns, Max Van Kleek, Michael Veale, Ulrik Lyngs, Jun Zhao, and Nigel Shadbolt. 2018. 'It's Reducing a Human Being to a Percentage': Perceptions of Justice in Algorithmic Decisions. In *Proceedings of the 2018 CHI Conference on Human Factors in Computing Systems*. ACM, 377.

[11] Anna Brown, Alexandra Chouldechova, Emily Putnam-Hornstein, Andrew Tobin, and Rhema Vaithianathan. 2019. Toward Algorithmic Accountability in Public Services: A Qualitative Study of Affected Community Perspectives on Algorithmic Decision-making in Child Welfare Services. In *Proceedings of the 2019 CHI Conference on Human Factors in Computing Systems, CHI 2019, Glasgow, Scotland, UK, May 04-09, 2019*. 41. https://doi.org/10.1145/3290605.3300271

[12] Ernest W. Burgess. 1928. Factors determining success or failure on parole. In *The workings of the indeterminate-sentence law and parole system in Illinois*, A. A. Bruce, A. J. Harno, E. W. Burgess, and J. Landesco (Eds.). Springfield, IL: State Board of Parole, 221–234.

[13] Ángel Alexander Cabrera, Will Epperson, Fred Hohman, Minsuk Kahng, Jamie Morgenstern, and Duen Horng Chau. 2019. FairVis: Visual analytics for discovering intersectional bias in machine learning. In *2019 IEEE Conference on Visual Analytics Science and Technology (VAST)*. IEEE, 46–56.

[14] Rich Caruana, Yin Lou, Johannes Gehrke, Paul Koch, Marc Sturm, and Noemie Elhadad. 2015. Intelligible models for healthcare: Predicting pneumonia risk and hospital 30-day readmission. In *Proceedings of the 21th ACM SIGKDD International Conference on Knowledge Discovery and Data Mining*. ACM, 1721–1730.

[15] Kathy Charmaz. 2014. *Constructing grounded theory*. sage.

[16] Alexandra Chouldechova. 2017. Fair prediction with disparate impact: A study of bias in recidivism prediction instruments. *Big data* 5, 2 (2017), 153–163.

[17] Sam Corbett-Davies and Sharad Goel. 2018. The Measure and Mismeasure of Fairness: A Critical Review of Fair Machine Learning. *CoRR* abs/1808.00023 (2018). arXiv:1808.00023 http://arxiv.org/abs/1808.00023

[18] Maria De-Arteaga, Riccardo Fogliato, and Alexandra Chouldechova. 2020. A Case for Humans-in-the-Loop: Decisions in the Presence of Erroneous Algorithmic Scores. In *Proceedings of the 2020 CHI Conference on Human Factors in Computing Systems*. 1–12.

[19] Sarah Dean, Mihaela Curmei, and Benjamin Recht. 2020. Designing Recommender Systems with Reachability in Mind. *Workshop on Participatory Approaches to Machine Learning* (2020), –.

[20] Matthew DeMichele, Peter Baumgartner, Michael Wenger, Kelle Barrick, Megan Comfort, and Shilpi Misra. 2018. The public safety assessment: A re-validation and assessment of predictive utility and differential prediction by race and gender in kentucky. *Available at SSRN 3168452* (2018).

[21] Sarah L Desmarais, Samantha A Zottola, Sarah E Duhart Clarke, and Evan M Lowder. 2020. Predictive Validity of Pretrial Risk Assessments: A Systematic Review of the Literature. *Criminal Justice and Behavior* (2020), 0093854820932959.

[22] Alan J Dettlaff. 2014. The evolving understanding of disproportionality and disparities in child welfare. In *Handbook of child maltreatment*. Springer, 149–168.

[23] Jonathan Dodge, Q. Vera Liao, Yunfeng Zhang, Rachel K. E. Bellamy, and Casey Dugan. 2019. Explaining Models: An Empirical Study of How Explanations Impact Fairness Judgment. In *Proceedings of the 24th International Conference on Intelligent User Interfaces* (Marina del Ray, California) *(IUI '19)*. ACM, New York, NY, USA, 275–285. https://doi.org/10.1145/3301275.3302310

[24] Cynthia Dwork, Moritz Hardt, Toniann Pitassi, Omer Reingold, and Richard S. Zemel. 2012. Fairness through awareness. In *Innovations in Theoretical Computer Science 2012, Cambridge, MA, USA, January 8-10, 2012*. 214–226. https://doi.org/10.1145/2090236.2090255

[25] Virginia Eubanks. 2018. *Automating inequality: How high-tech tools profile, police, and punish the poor.* St. Martin's Press.

[26] Sorelle A Friedler, Carlos Scheidegger, and Suresh Venkatasubramanian. 2016. On the (im) possibility of fairness. *arXiv preprint arXiv:1609.07236* (2016).

[27] Jeremy D Goldhaber-Fiebert and Lea Prince. 2019. Impact Evaluation of a Predictive Risk Modeling Tool for Allegheny County's Child Welfare Office. *Pittsburgh: Allegheny County* (2019).

[28] Nina Grgic-Hlaca, Elissa M. Redmiles, Krishna P. Gummadi, and Adrian Weller. 2018. Human Perceptions of Fairness in Algorithmic Decision Making: A Case Study of Criminal Risk Prediction. In *Proceedings of the 2018 World Wide Web Conference* (Lyon, France) *(WWW '18)*. International World Wide Web Conferences Steering Committee, 903–912. https://doi.org/10.1145/3178876.3186138

[29] Nina Grgić-Hlača, Muhammad Bilal Zafar, Krishna P. Gummadi, and Adrian Weller. 2018. Beyond distributive fairness in algorithmic decision making: Feature selection for procedurally fair learning. In *Proceedings of the 32nd AAAI Conference on Artificial Intelligence*.

[30] Moritz Hardt, Eric Price, and Nati Srebro. 2016. Equality of Opportunity in Supervised Learning. In *Advances in Neural Information Processing Systems 29: Annual Conference on Neural Information Processing Systems 2016, December 5-10, 2016, Barcelona, Spain*. 3315–3323. http://papers.nips.cc/paper/6374-equality-of-opportunity-in-supervised-learning

[31] Úrsula Hébert-Johnson, Michael P. Kim, Omer Reingold, and Guy N. Rothblum. 2018. Multicalibration: Calibration for the (Computationally-Identifiable) Masses. In *Proceedings of the 35th International Conference on Machine Learning, ICML 2018, Stockholmsmässan, Stockholm, Sweden, July 10-15, 2018*. 1944–1953. http://proceedings.mlr.press/v80/hebert-johnson18a.html

[32] Kenneth Holstein, Jennifer Wortman Vaughan, Hal Daumé III, Miroslav Dudík, and Hanna M. Wallach. 2019. Improving Fairness in Machine Learning Systems: What Do Industry Practitioners Need?. In *Proceedings of the 2019 CHI Conference on Human Factors in Computing Systems, CHI 2019, Glasgow, Scotland, UK, May 04-09, 2019*. 600. https://doi.org/10.1145/3290605.3300830

[33] Christina Ilvento. 2020. Metric Learning for Individual Fairness. In *1st Symposium on Foundations of Responsible Computing, FORC 2020, June 1-3, 2020, Harvard University, Cambridge, MA, USA (virtual conference) (LIPIcs, Vol. 156)*, Aaron Roth (Ed.). Schloss Dagstuhl - Leibniz-Zentrum für Informatik, 2:1–2:11. https://doi.org/10.4230/LIPIcs.FORC.2020.2

[34] Abby Everett Jaques. 2019. Why the Moral Machine Is a Monster. In *University of Miami Law School: We Robot Conference*. https://robots.law.miami.edu/2019/wp-content/uploads/2019/03/MoralMachineMonster.pdf

[35] Will Johnson. 2004. Effectiveness of California's child welfare structured decision making (SDM) model: a prospective study of the validity of the California Family Risk Assessment. *Madison (Wisconsin, USA): Children's Research Center* (2004).

[36] Caroline M Johnston, Simon Blessenohl, and Phebe Vayanos. [n.d.]. Preference Elicitation and Aggregation to Aid with Patient Triage during the COVID-19 Pandemic. *Workshop on Participatory Approaches to Machine Learning* ([n. d.]).

[37] Matthew Joseph, Michael J. Kearns, Jamie H. Morgenstern, and Aaron Roth. 2016. Fairness in Learning: Classic and Contextual Bandits. In *Advances in Neural Information Processing Systems 29: Annual Conference on Neural Information Processing Systems 2016, December 5-10, 2016, Barcelona, Spain*. 325–333. http://papers.nips.cc/paper/6355-fairness-in-learning-classic-and-contextual-bandits

[38] Christopher Jung, Michael J. Kearns, Seth Neel, Aaron Roth, Logan Stapleton, and Zhiwei Steven Wu. 2019. Eliciting and Enforcing Subjective Individual Fairness. *CoRR* abs/1905.10660 (2019). arXiv:1905.10660 http://arxiv.org/abs/1905.10660

[39] Christopher Jung, Michael J. Kearns, Seth Neel, Aaron Roth, Logan Stapleton, and Zhiwei Steven Wu. 2019. Eliciting and Enforcing Subjective Individual Fairness. *CoRR* abs/1905.10660 (2019). arXiv:1905.10660 http://arxiv.org/abs/1905.10660

[40] Anson Kahng, Min Kyung Lee, Ritesh Noothigattu, Ariel D. Procaccia, and Christos-Alexandros Psomas. 2019. Statistical Foundations of Virtual Democracy. In *Proceedings of the 36th International Conference on Machine Learning, ICML 2019, 9-15 June 2019, Long Beach, California, USA*. 3173–3182. http://proceedings.mlr.press/v97/kahng19a.html

[41] Faisal Kamiran and Toon Calders. 2012. Data preprocessing techniques for classification without discrimination. *Knowledge and Information Systems* 33, 1 (2012), 1–33.
[42] Michael J. Kearns, Seth Neel, Aaron Roth, and Zhiwei Steven Wu. 2018. Preventing Fairness Gerrymandering: Auditing and Learning for Subgroup Fairness. In *Proceedings of the 35th International Conference on Machine Learning, ICML 2018, Stockholmsmässan, Stockholm, Sweden, July 10-15, 2018.* 2569–2577. http://proceedings.mlr.press/v80/kearns18a.html
[43] Danielle Leah Kehl and Samuel Ari Kessler. 2017. Algorithms in the criminal justice system: Assessing the use of risk assessments in sentencing. (2017).
[44] Amir E Khandani, Adlar J Kim, and Andrew W Lo. 2010. Consumer credit-risk models via machine-learning algorithms. *Journal of Banking & Finance* 34, 11 (2010), 2767–2787.
[45] Jon Kleinberg, Himabindu Lakkaraju, Jure Leskovec, Jens Ludwig, and Sendhil Mullainathan. 2017. Human decisions and machine predictions. *The quarterly journal of economics* 133, 1 (2017), 237–293.
[46] Jon M. Kleinberg, Sendhil Mullainathan, and Manish Raghavan. 2017. Inherent Trade-Offs in the Fair Determination of Risk Scores. In *8th Innovations in Theoretical Computer Science Conference, ITCS*.
[47] Felicitas Kraemer, Kees van Overveld, and Martin Peterson. 2011. Is there an ethics of algorithms? *Ethics and Information Technology* 13 (2011), 251—-260.
[48] Vivian Lai and Chenhao Tan. 2019. On human predictions with explanations and predictions of machine learning models: A case study on deception detection. In *Proceedings of the Conference on Fairness, Accountability, and Transparency.* 29–38.
[49] Min Kyung Lee. 2018. Understanding perception of algorithmic decisions: Fairness, trust, and emotion in response to algorithmic management. *Big Data & Society* 5, 1 (2018), 2053951718756684.
[50] Min Kyung Lee and Su Baykal. 2017. Algorithmic mediation in group decisions: Fairness perceptions of algorithmically mediated vs. discussion-based social division. In *Proceedings of the 2017 ACM Conference on Computer Supported Cooperative Work and Social Computing.* ACM, 1035–1048.
[51] Min Kyung Lee, Anuraag Jain, Hea Jin Cha, Shashank Ojha, and Daniel Kusbit. 2019. Procedural justice in algorithmic fairness: Leveraging transparency and outcome control for fair algorithmic mediation. *Proceedings of the ACM on Human-Computer Interaction* 3, CSCW (2019), 1–26.
[52] Min Kyung Lee, Ji Tae Kim, and Leah Lizarondo. 2017. A human-centered approach to algorithmic services: Considerations for fair and motivating smart community service management that allocates donations to non-profit organizations. In *Proceedings of the 2017 CHI Conference on Human Factors in Computing Systems.* ACM, 3365–3376.
[53] Min Kyung Lee, Daniel Kusbit, Anson Kahng, Ji Seong Tae, Xinran Yuan, A. D. C. Chan, Ritesh Noothigattu, Daniel See, Siheon Lee, Christos-Alexandros Psomas, and Ariel D. Procaccia. 2018. WeBuildAI : Participatory Framework for Fair and Efficient Algorithmic Governance.
[54] David B Marshall and Diana J English. 2000. Neural network modeling of risk assessment in child protective services. *Psychological Methods* 5, 1 (2000), 102.
[55] John Monahan. 2017. Risk assessment in sentencing. *Academy for Justice, a Report on Scholarship and Criminal Justice Reform (Erik Luna ed., 2017, Forthcoming)* (2017).
[56] John Monahan, Anne Metz, and Brandon L Garrett. 2018. Judicial Appraisals of Risk Assessment in Sentencing. (2018).
[57] Arvind Narayanan. 2018. Translation tutorial: 21 fairness definitions and their politics. In *Proc. Conf. Fairness Accountability Transp., New York, USA*.
[58] Jakob Nielsen. 1994. *Usability engineering.* Morgan Kaufmann.
[59] Dino Pedreshi, Salvatore Ruggieri, and Franco Turini. 2008. Discrimination-aware data mining. In *Proceedings of the 14th ACM SIGKDD international conference on knowledge discovery and data mining.* ACM, 560–568.
[60] Geoff Pleiss, Manish Raghavan, Felix Wu, Jon M. Kleinberg, and Kilian Q. Weinberger. 2017. On Fairness and Calibration. In *Advances in Neural Information Processing Systems 30: Annual Conference on Neural Information Processing Systems 2017, 4-9 December 2017, Long Beach, CA, USA.* 5684–5693. http://papers.nips.cc/paper/7151-on-fairness-and-calibration
[61] Edmond Awad Sohan Dsouza Iyad Rahwan Pradeep Ravikumar Ritesh Noothigattu, Snehalkumar 'Neil' S. Gaikwad and Ariel D. Procaccia. 2018. A voting-based system for ethical decision making. In *Proceedings of the 32nd AAAI Conference on Artificial Intelligence (AAAI)*.
[62] Samantha Robertson and Niloufar Salehi. 2020. What If I Don't Like Any Of The Choices? The Limits of Preference Elicitation for Participatory Algorithm Design. *arXiv preprint arXiv:2007.06718* (2020).
[63] Debjani Saha, Candice Schumann, Duncan C. McElfresh, John P. Dickerson, Michelle L. Mazurek, and Michael Carl Tschantz. 2020. Measuring Non-Expert Comprehension of Machine Learning Fairness Metrics. In *Proceedings of the 37th International Conference on Machine Learning, ICML 2020, Vienna, Austria, July 12–18, 2020*.
[64] Nripsuta Ani Saxena, Karen Huang, Evan DeFilippis, Goran Radanovic, David C. Parkes, and Yang Liu. 2019. How Do Fairness Definitions Fare?: Examining Public Attitudes Towards Algorithmic Definitions of Fairness. In *Proceedings of the 2019 AAAI/ACM Conference on AI, Ethics, and Society.* ACM, 99–106.
[65] Nicholas Scurich and John Monahan. 2016. Evidence-based sentencing: Public openness and opposition to using gender, age, and race as risk factors for recidivism. *Law and Human Behavior* 40, 1 (2016), 36.
[66] Hetan Shah. 2018. Algorithmic accountability. *Philosophical Transactions of the Royal Society A: Mathematical, Physical and Engineering Sciences* 376, 2128 (2018), 20170362.
[67] Vernon C Smith, Adam Lange, and Daniel R Huston. 2012. Predictive modeling to forecast student outcomes and drive effective interventions in online community college courses. *Journal of Asynchronous Learning Networks* 16, 3 (2012), 51–61.
[68] Megha Srivastava, Hoda Heidari, and Andreas Krause. 2019. Mathematical notions vs. human perception of fairness: A descriptive approach to fairness for machine learning. In *Proceedings of the 25th ACM SIGKDD International Conference on Knowledge Discovery & Data Mining.* 2459–2468.
[69] Rhema Vaithianathan, Nan Jiang, Tim Maloney, Parma Nand, and Emily Putnam-Hornstein. 2017. Developing Predictive Risk Models to Support Child Maltreatment Hotline Screening Decisions. https://www.alleghenycountyanalytics.us/wp-content/uploads/2017/04/Developing-Predictive-Risk-Models-package-with-cover-1-to-post-1.pdf
[70] Michael Veale, Max Van Kleek, and Reuben Binns. 2018. Fairness and accountability design needs for algorithmic support in high-stakes public sector decision-making. In *Proceedings of the 2018 chi conference on human factors in computing systems.* 1–14.
[71] Sahil Verma and Julia Rubin. 2018. Fairness definitions explained. In *2018 IEEE/ACM International Workshop on Software Fairness (FairWare).* IEEE, 1–7.
[72] AJ Wang. 2018. Procedural Justice and Risk-Assessment Algorithms. (2018).
[73] Jiaxuan Wang, Jeeheh Oh, Haozhu Wang, and Jenna Wiens. 2018. Learning credible models. In *Proceedings of the 24th ACM SIGKDD International Conference on Knowledge Discovery & Data Mining.* 2417–2426.
[74] Ruotong Wang, F Maxwell Harper, and Haiyi Zhu. 2020. Factors Influencing Perceived Fairness in Algorithmic Decision-Making: Algorithm Outcomes, Development Procedures, and Individual Differences. In *Proceedings of the 2020 CHI Conference on Human Factors in Computing Systems.* 1–14.
[75] James Wexler, Mahima Pushkarna, Tolga Bolukbasi, Martin Wattenberg, Fernanda Viégas, and Jimbo Wilson. 2019. The what-if tool: Interactive probing of machine learning models. *IEEE transactions on visualization and computer graphics* 26, 1 (2019), 56–65.
[76] Meredith Whittaker, Kate Crawford, Roel Dobbe, Genevieve Fried, Elizabeth Kaziunas, Varoon Mathur, Sarah Mysers West, Rashida Richardson, Jason Schultz, and Oscar Schwartz. 2018. *AI now report 2018.* AI Now Institute at New York University.
[77] Pak-Hang Wong. 2020. Democratizing Algorithmic Fairness. *Philosophy & Technology* 33 (2020), 225–244. https://doi.org/10.1145/3290605.3300830
[78] Allison Woodruff, Sarah E Fox, Steven Rousso-Schindler, and Jeffrey Warshaw. 2018. A qualitative exploration of perceptions of algorithmic fairness. In *Proceedings of the 2018 CHI Conference on Human Factors in Computing Systems.* ACM, 656.
[79] Bowen Yu, Ye Yuan, Loren Terveen, Zhiwei Steven Wu, Jodi Forlizzi, and Haiyi Zhu. 2020. Keeping Designers in the Loop: Communicating Inherent Algorithmic Trade-offs Across Multiple Objectives. In *Proceedings of the 2020 ACM Designing Interactive Systems Conference.* 1245–1257.
[80] Muhammad Bilal Zafar, Isabel Valera, Manuel Gomez-Rodriguez, and Krishna P. Gummadi. 2017. Fairness Beyond Disparate Treatment & Disparate Impact: Learning Classification without Disparate Mistreatment. In *Proceedings of the 26th International Conference on World Wide Web, WWW.* ACM, 1171–1180.
[81] Haiyi Zhu, Bowen Yu, Aaron Halfaker, and Loren Terveen. 2018. Value-sensitive algorithm design: Method, case study, and lessons. *Proceedings of the ACM on Human-Computer Interaction* 2, CSCW (2018), 194.